# Collapse-revival of quantum discord and entanglement

Xue-Qun Yan[†] and Bo-Ying Zhang

Department of Physics, School of Science, Tianjin Polytechnic University, Tianjin 300387, China

Tianjin Key Laboratory of Advanced Technology of Electrical Engineering and Energy, Tianjin 300387, China

In this paper the correlations dynamics of two atoms in the case of a micromaser-type system is investigated. Our results predict certain quasi-periodic collapse and revival phenomena for quantum discord and entanglement when the field is in Fock state and the two atoms are initially in maximally mixed state, which is a unique separable state. Our calculations also show that the oscillations of the time evolution of both quantum discord and entanglement are almost in phase and they both have similar evolution behavior in some time range. The fact reveals the consistency of quantum discord and entanglement in some dynamical aspects.

*Introduction*. The existence of nonclassical correlations between distinct quantum systems is one of the fundamental differences between quantum and classical physics. Entanglement, which is a key resource in quantum information science, has been considered to be the only kind of nonclassical correlation. However, it has become clear by now that entanglement does not include all nonclassical correlations although most of the quantum computation and communication processes rely on it. The nonclassical correlation, which cannot be captured by entanglement measure, is called quantum discord [1]. Both entanglement and quantum discord have become a fundamental resource in quantum information processing and a number of theoretical studies have been performed on this subject in recent years [2]. Unfortunately, we do not have yet a closed theory for the description of the nonclassical correlations, and there is no clear evidence of the relation between quantum discord and entanglement [3], in spite of some calculations of quantum discord for several families of quantum states and comparison with the entanglement have been presented in the literature [4-6]. Recently some studies have also showed that quantum discord is a resource in the tasks of entanglement distribution [7,8], and an operational interpretation of quantum discord has been given that quantum discord is a quantitative measure about the performance in the quantum-state merging [9,10]. More recently, in Ref. [11], Streltsov and Zurek showed that if a measurement device is in a nonclassical state, the measurement results cannot be communicated perfectly by classical means. In this case some part of the information in the measurement apparatus is lost in the process of communication, and the amount of this lost information turns out to be the quantum discord. They also found that the information loss occurs even when the apparatus is not entangled with the system of interest. In addition, the study of nature of correlations in a quantum state has also attracted the impressive amount of attention and efforts for both of theoreticians and of experimentalists. There are many articles devoted to the investigations of this aspect [12-28]. A prevailing observation in all results obtained is that the discord is directly associated to non-trivial properties of states. Particularly, Knill and Laflamma showed that quantum computation in which only one qubit is in a nonmaximally mixed state, while the rest are in maximally mixed state, can achieve an exponential improvement in efficiency over classical computers for a limited set of tasks [29].

---

[†] E-mail: xqyan867@tom.com

The studies of quantum correlations in different dynamical systems are another important topic not only from a fundamental point of view, but also for practical purposes [2,30,31]. For the nonce, many interesting phenomena of correlations dynamics have already been discovered, including entanglement sudden death [32] and sudden change of quantum discord [33], freezing of quantum correlations [27] and recovery of quantum correlations in absence of system-environment back-action [30,34,35]. In the paper we will consider a pair of two-level atoms going though a cavity one after another. A detailed understanding of the evolution of entanglement and quantum discord in the system is valuable for both fundamental theoretical investigations and experimentally realizable systems. We study the dynamics of a pair of such atoms numerically and predict certain quasi-periodic collapse and revival phenomena for quantum correlations when the field is in Fock state and the two atoms are initially in a unique separable state, that is, the maximally mixed state.

*Dynamics of the two atoms in a micromaser-type system.* Let us consider a micromaser-type system in which atoms are injected at a rate low enough that at most one atom at a time is inside the cavity, and the time of flight through the cavity $t$ is the same for every atom. For simplicity we will suppose that the cavity is of a non-leaky type. The cavity-QED experiments are, in fact, very close to such situation [36-39]. While an atom flies through the cavity, the coupled atom-field system is described by the Jaynes-Cummings Hamiltonian. The interaction Hamiltonian in a rotating frame at the cavity mode frequency and in the rotating wave approximation, at exact resonance, can be written as $H_I = g(\sigma_+ a + a^+ \sigma_-)$. Where $a(a^+)$ denote the annihilation (creation) operators of the single-mode cavity field and $\sigma_+(\sigma_-)$ represent the raising (lowing) operators of the atom, and $g$ is the atom-field coupling constant.

We assume that the cavity field is initially in the $n-$ photon Fock state $|n\rangle$ and two spatially separated two-level atoms $A$ and $B$ are initially prepared in a state $\rho_{AB}(0)$. The two atoms pass through the cavity one after the other and have no direct interaction. We further assume that initially the total system is found in product state $\rho(0) = \rho_{AB}(0) \otimes |n\rangle\langle n|$. In the standard basis $|11\rangle$, $|10\rangle$, $|01\rangle$ and $|00\rangle$, if the initial state takes the form

$$\rho_{AB}(0) = \begin{pmatrix} \rho_{11}(0) & 0 & 0 & 0 \\ 0 & \rho_{22}(0) & \rho_{23}(0) & 0 \\ 0 & \rho_{32}(0) & \rho_{33}(0) & 0 \\ 0 & 0 & 0 & \rho_{44}(0) \end{pmatrix}, \quad (1)$$

the form remains under evolution.

For the initial state of Eq. (1), the matrix element dynamics of the reduced density matrix $\rho_{AB}(t)$ of the two atoms after passing though the cavity field is given by

$$\rho_{11}(t) = \rho_{11}(0)c_{n+1}^4 + \rho_{22}(0)s_n^2 c_{n+1}^2 + \rho_{33}(0)s_n^2 c_n^2 + \rho_{44}(0)s_n^2 s_{n-1}^2$$
$$+ (\rho_{23}(0) + \rho_{32}(0))s_n^2 c_{n+1} c_n,$$

$$\rho_{22}(t) = \rho_{11}(0)s_{n+1}^2 c_{n+1}^2 + \rho_{22}(0)c_n^2 c_{n+1}^2 + \rho_{33}(0)s_n^4 + \rho_{44}(0)s_n^2 c_{n-1}^2$$
$$- (\rho_{23}(0) + \rho_{32}(0))s_n^2 c_{n+1} c_n,$$

$$\rho_{33}(t) = \rho_{11}(0)s_{n+1}^2 c_{n+2}^2 + \rho_{22}(0)s_{n+1}^4 + \rho_{33}(0)c_{n+1}^2 c_n^2 + \rho_{44}(0)s_n^2 c_n^2$$
$$- (\rho_{23}(0) + \rho_{32}(0))s_{n+1}^2 c_{n+1} c_n,$$

$$\rho_{44}(t) = \rho_{11}(0)s_{n+1}^2 s_{n+2}^2 + \rho_{22}(0)s_{n+1}^2 c_{n+1}^2 + \rho_{33}(0)s_{n+1}^2 c_n^2 + \rho_{44}(0)c_n^4$$

$$+ \left(\rho_{23}(0) + \rho_{32}(0)\right) s_n^2 c_{n+1} c_n,$$

$$\rho_{23}(t) = \rho_{11}(0)s_{n+1}^2 c_{n+1} c_{n+2} - \rho_{22}(0)s_{n+1}^2 c_{n+1} c_n - \rho_{33}(0)s_n^2 c_{n+1} c_n$$

$$+ \rho_{44}(0)s_n^2 c_n c_{n-1} + \rho_{23}(0)c_{n+1}^2 c_n^2 + \rho_{32}(0)s_{n+1}^2 s_n^2,$$

$$\rho_{32}(t) = \rho_{11}(0)s_{n+1}^2 c_{n+1} c_{n+2} - \rho_{22}(0)s_{n+2}^2 c_{n+1} c_n - \rho_{33}(0)s_n^2 c_{n+1} c_n$$

$$+ \rho_{44}(0)s_n^2 c_n c_{n-1} + \rho_{23}(0)s_{n+1}^2 s_n^2 + \rho_{32}(0)c_{n+1}^2 c_n^2. \tag{2}$$

Here $c_n = \cos\left(\sqrt{n}gt\right)$ and $s_n = \sin\left(\sqrt{n}gt\right)$. The solutions (2) are used to find the dynamics evolution of quantum correlations.

*Measures of quantum correlations.* In order to describe the dynamics evolution of entanglement, we use the concurrence [40], which for a mixed state of two qubits is given by $C(\rho) = \max(0, \sqrt{\lambda_1} - \sqrt{\lambda_2} - \sqrt{\lambda_3} - \sqrt{\lambda_4})$, where $\lambda_i$ are the eigenvalues (in descending order) of $\rho(\sigma_y \otimes \sigma_y)\rho^*(\sigma_y \otimes \sigma_y)$, in which $\rho$ is the density matrix of the system. The concurrence varies between 0 when qubits are separable and 1 when they are maximally entangled. In the case of the density matrix with the matrix elements in Eq. (2), the concurrence $C$ can be easily obtained as

$$C = 2\max\left\{0, |\rho_{23}(t)| - \sqrt{\rho_{11}(t)\rho_{44}(t)}\right\}. \tag{3}$$

Moreover, for a given bipartite state $\rho_{AB}$ shared by two parties $A$ and $B$, the quantum discord is usually defined as the difference between the quantum mutual information and the classical correlation,

$$D(\rho_{AB}) = I(\rho_{AB}) - C'(\rho_{AB}). \tag{4}$$

Here, the quantum mutual information of two subsystems is given by

$$I(\rho_{AB}) = S(\rho_A) + S(\rho_B) - S(\rho_{AB}), \tag{5}$$

where $S(\rho_j) = -Tr_j(\rho_j \log_2 \rho_j) = -\sum_i \lambda_j^i \log_2 \lambda_j^i$ is the von Neumann entropy with $\{\lambda_j^i\}$ being the nonzero eigenvalues of the quantum state $\rho_j$, and the subscript $j$ indicates either the subsystem $A(B)$ or the total system. Following [41], the amount of classical correlations in $\rho_{AB}$ can be quantified by

$$C'(\rho_{AB}) = S(\rho_A) - \min_{\{\Pi_k\}}[S(\rho_{AB}|\{\Pi_k\})], \qquad (6)$$

where the minimum is taken over the complete set of projective measurements $\{\Pi_k\}$ and $S(\rho_{AB}|\{\Pi_k\}) = \sum_k p_k S(\rho_k)$ is the quantum conditional entropy of $A$ given the complete measurements on subsystem $B$, with $\rho_k = Tr_B(\Pi_k \rho_{AB} \Pi_k)/p_k$ and $p_k = Tr_{AB}(\rho_{AB} \Pi_k)$. Substituting Eqs. (5) and (6) into (4), we obtain

$$D(\rho_{AB}) = S(\rho_B) - S(\rho_{AB}) + \min_{\{\Pi_k\}} S(\rho_{AB}|\{\Pi_k\}). \qquad (7)$$

It is worth pointing out that completely quantifying quantum discord in the most general cases is still an open problem, but for certain special class of states analytical results are available [5,6].

To calculate quantum discord we use the algorithm introduced by Ali *et al* [5], which is an extension of the method of Luo [6]. Despite some counterexamples have been given in [42,43], Huang confirmed numerically, in a recent article [44], that the Ali *et al* algorithm is valid with worst-case absolute error $0.0021$ for two-qubit X states. Our calculations show that this error will not affect our conclusions.

We first obtain the eigenvalues of the matrix $\rho_{AB}(t)$

$$\lambda_{0,1} = \frac{1}{2}[(\rho_{11}(t) + \rho_{44}(t)) \pm |\rho_{11}(t) - \rho_{44}(t)|],$$
$$\lambda_{2,3} = \frac{1}{2}[(\rho_{22}(t) + \rho_{33}(t)) \pm \sqrt{(\rho_{22}(t) - \rho_{33}(t))^2 + 4|\rho_{23}(t)|^2}]. \qquad (8)$$

With Eq. (8) in hand, one can immediately get the entropy of the total system $S(\rho_{AB})$. In addition, we can express the von Neumann entropy of the reduced density matrix $\rho_B$ in terms of Eq. (2) as follows

$$S(\rho_B) = -[(\rho_{11}(t) + \rho_{33}(t))\log_2(\rho_{11}(t) + \rho_{33}(t))$$
$$+ (\rho_{22}(t) + \rho_{44}(t))\log_2(\rho_{22}(t) + \rho_{44}(t))]. \qquad (9)$$

Based on the above discussion and using the algorithm of [5], the quantum discord are given by

$$D(\rho_{AB}) = S(\rho_B) - S(\rho_{AB}) + \min\left\{S(\rho_0)\big|_{\theta_1}, S(\rho_{AB}|\{\Pi_k\})\big|_{\theta_2,\theta_3}\right\}, \qquad (10)$$

where

$$S(\rho_0)\big|_{\theta_1} = h(x_1), S(\rho_{AB}|\{\Pi_k\})\big|_{\theta_2,\theta_3} = (\rho_{22}(t)+\rho_{44}(t))h(x_2) + (\rho_{11}(t)+\rho_{33}(t))h(x_3). \quad (11)$$

Here $h(x) = -x\log x - (1-x)\log(1-x)$ is the binary entropy function and $x_i = \dfrac{1+\theta_i}{2}$, $i=1,2,3$, where $\theta_1 = \sqrt{(2\rho_{11}(t)+2\rho_{22}(t)-1)^2 + 4|\rho_{23}(t)|^2}$, $\theta_2 = \dfrac{|\rho_{22}(t)-\rho_{44}(t)|}{\rho_{22}(t)+\rho_{44}(t)}$ and

$$\theta_3 = \frac{|\rho_{11}(t)-\rho_{33}(t)|}{\rho_{11}(t)+\rho_{33}(t)}.$$

Having obtained the results above, we are therefore in a position to discuss the dynamics behavior of the quantum discord and entanglement related to the system considered. This will be done in the following section.

*Numerical results and discussions.* Our goal in this article is to clarify what peculiar dynamics happens for quantum correlations of the system under consideration. To make our results concrete, we restrict our analysis to the initial Werner state $\rho_W(0) = r|\psi^+\rangle\langle\psi^+| + \dfrac{(1-r)}{4}I_4$. Here $|\psi^+\rangle = (|10\rangle + |01\rangle)/\sqrt{2}$, and $I_4$ is the $4\times 4$ identity matrix. $r$ varies from 0 to 1, and for $r=0$ the Werner state become maximally mixed states, while for $r=1$ they are the well-known Bell states.

The dynamical evolution of the concurrence and quantum discord are shown in Figs. 1-3 as a function of the Rabi angle $gt$ ($t$ is the time spent by the atom inside the cavity) for different initial states. It is seen from these figures that the oscillations of time evolution of both entanglement and quantum discord are almost in phase in some time range, in particular the numerical value of the quantum discord remains close to the concurrence for most of cases, in spite of concurrence and quantum discord are different measures that are based on different concepts and use different mathematical entities. The fact demonstrates the consistency of quantum discord and entanglement in some dynamical aspects.

Figure 1 shows the time evolution of the entanglement and quantum discord for the initial maximally mixed state of two-qubit $I_4/4$ with the Fock state field $n=0$ case (i.e. the ground state $|0\rangle$). It is seen that both entanglement and quantum discord are zero at earlier times, however, after some finite period of time the correlations start to build up via atom-photon interaction inside the cavity, although no single atom interacts directly with another. One should note that the quantum discord show up before the entanglement is generated. It is similar to the case of [45], in where quantum discord show up as precursors of entanglement when two initially excited qubits interact with each other through a common environment and also through dipole forces. From the figure we can see that in the evolution, concurrence abruptly becomes zero and remains zero for some time interval, but quantum discord is always continuous change. Similar behavior has been reported in different cases, both theoretically and experimentally (see, for example, [2] and the references therein). On the other hand, we can also see that, for some time interval, concurrence is zero while discord exist extreme values. Maybe it means that quantum discord is more robust than entanglement.

It is remarkable to see in Fig. 2 that for the initial maximally mixed state of two atoms and higher values of the field photons, there are certain quasi-periodic collapse and revival phenomena of the correlations oscillations. From the figure, firstly one can see the envelope of the correlations oscillations 'collapses' to nearly zero. However, as time increases we encounter a 'revival' of the collapsed correlations, and there are only small fluctuations between two envelopes. The behavior of collapse and revival of correlations is repeated and the amplitude of the envelope of the oscillations slightly decreases as time increases. Moreover, the fluctuations between two envelopes increase with time increase. By comparing Fig. 2(a) with Fig. 2(b) we see that as $n$ increases, the periodic of the oscillations become larger, while the fluctuations between two envelopes decrease.

In order to see effects of initial entangled states on the evolution of both quantum discord and entanglement, we examine the dynamical evolution of the quantum discord and entanglement for the initial entangled Werner state, i.e., $\rho_W(0) = r|\psi^+\rangle\langle\psi^+| + \frac{(1-r)}{4}I_4$, $r \neq 0$ (see Fig. 3, where we set $r = 0.2$ and $n = 10$). It is obvious that at earlier times both quantum discord and entanglement are no longer zero and either entanglement or quantum discord become larger than that for the case of $r = 0$. Also, it is found that the phenomenon of collapse and revival is sensitive to the initial atoms states and field states. Our results seem to suggest that the periodic exchange of energy between the atomic and the field oscillators leads to the collapse and revival of quantum discord and entanglement.

The above analysis can easily be extended to study correlations dynamics starting from different initial conditions. The details of the evolution for this case will be considered elsewhere.

*Conclusions*. We have investigated the correlations dynamics of two atoms in the case of a micromaser-type system and predicted an interesting quasi-periodic collapse and revival phenomenon for quantum discord and entanglement for the field initially in the Fock state and the atoms initially in a unique separable state, that is, the maximum mixed state, and the phenomenon is sensitive to the initial atoms and field states. In addition, we have also observed that in some time range concurrence and quantum discord have similar evolution behavior, despite concurrence and quantum discord are different measures for quantum correlations. It shows the consistency of quantum discord and entanglement in some dynamical aspects. Although the theoretical interpretation of our main results is not exactly clear, we believe that our results greatly enrich our knowledge and understanding of the dynamics of quantum correlations, and are significant for clarifying the connection between quantum discord and entanglement.

## Acknowledgement

We are grateful to Yu-Chun Wu for careful reading of the original manuscript.

## References

[1] H. Olliver and W. H. Zurek, Phys. Rev. Lett. **88**, 017901 (2001).

[2] K. Modi, A. Brodutch, H. Cable, T. Paterek, and V. Vedral, Rev. Mod. Phys. **84**，1655 (2012).

[3] L. C. Céleri, J. Maziero, and R. M. Serra, Int. J. Quantum Inform. **09**, 1837 (2011).

[4] R. Tanaś, Phys. Scr. **T153**, 014059 (2013).

[5] M. Ali, A. R. P. Rau, and G. Alber, Phys. Rev. A **81**, 042105 (2010).

[6] S. Luo, Phys. Rev. A **77**, 042303 (2008).

[7] A. Streltsov, H. Kampermann, and D. Bruß, Phys. Rev. Lett. **108**, 250501 (2012).

[8] T. K. Chuan, J. Maillard, K. Modi, T. Paterek, M. Paternostro, and M. Piani, Phys. Rev. Lett. **109**, 070501 (2012).

[9] D. Cavalcanti, L. Aolita, S. Boixo, K. Modi, M. Piani, and A. Winter, Phys Rev A **83**, 032324

(2011).

[10] V. Madhok and A. Datta, Phys Rev A **83**, 032323 (2011).

[11] A. Streltsov and W. H. Zurek, Phys. Rev. Lett. **111**, 040401 (2013).

[12] T. Werlang, S. Souza, F. F. Fanchini, and C. J. Villas-Boas, Phys. Rev. A **80**, 024103 (2009).

[13] J. Maziero, L. C. Céleri, R. M. Serra, V. Vedral, Phys. Rev. A **80**, 044102 (2009).

[14] R. H. Xiao, Z. Y. Guo, S. Q. Zhu, and J. X. Fang, Int. J. Theor. Phys. **52**, 1721 (2013).

[15] R. Vasile, P. Giorda, S. Olivares, M. G. A. Paris, and S. Maniscalco, Phys. Rev. A **82**, 012313 (2010).

[16] A. Auyuanet and L. Davidovich, Phys. Rev. A **82**, 032112 (2010).

[17] B. Wang, Z. Y. Xu, Z. Q. Chen, and M. Feng, Phys. Rev. A **81**, 014101 (2010).

[18] G. Karpat and I. Gedik, Phys. Scr. **T153**, 014036 (2013).

[19] X. Q. Yan and Z. L. Yue, Chaos Solitons & Fractals **57**,117 (2013).

[20] B. P. Lanyon, P. Jurcevic, C. Hempel, M. Gessner, V. Vedral, R. Blatt, and C. F. Roos, Phys. Rev. Lett. **111**, 100504 (2013).

[21] J. P. G.. Pinto, G. Karpat, and F. F. Fanchini, Phys. Rev. A **88**, 034304 (2013).

[22] G. L. Giorgi, Phys. Rev. A **88**, 022315 (2013).

[23] M.-L. Hu, H. Fan, Ann. Phys. (NY) **327**, 851 (2012).

[24] M.-L. Hu, H. Fan, Ann. Phys. (NY) **327**, 2343 (2012).

[25] M.-L. Hu, D.-P. Tian, Ann. Phys. (NY) **343**, 132 (2014).

[26] B. Aaronson, R. Lo Franco, G. Compagno, and G. Adesso, New J. Phys. **15**, 093022 (2013).

[27] B. Aaronson, R. Lo Franco, and G. Adesso, Phys. Rev. A **88**, 012120 (2013).

[28] B. Bellomo, G. L. Giorgi, F. Galve, R. Lo Franco, G. Compagno, R. Zambrini, Phys. Rev. A **85**, 032104 (2012).

[29] E. Knill and R. Laflamme, Phys. Rev. Lett. **81,** 5672 (1998).

[30] J.-S. Xu, K. Sun, C.-F. Li, X.-Y. Xu, G.-C. Guo, E. Andersson, R. Lo Franco, and G. Compagno, Nature Commun. **4**, 2851 (2013).

[31] R. Lo Franco, B. Bellomo, S. Maniscalco, and G. Compagno, Int. J. Mod. Phys. B **27**, 1345053 (2013).

[32] T. Yu and J. H. Eberly, Phys. Rev. Lett. **93**, 140404 (2004).

[33] L. Mazzola, J. Piilo, and S. Maniscalco, Phys. Rev. Lett. **104**, 200401 (2010).

[34] A. D'Arrigo, R. Lo Franco, G. Benenti, E. Paladino, and G. Falci, Physica Scripta **T153**, 014014 (2013).

[35] R. Lo Franco, B. Bellomo, E. Andersson, and G. Compagno, Phys. Rev. A **85**, 032318 (2012).

[36] J. M. Raimond, M. Brune, and S. Haroche, Rev. Mod. Phys. **73**, 565 (2001).

[37] R. Lo Franco, G. Compagno, A. Messina, and A. Napoli, Phys. Rev. A **76**, 011804(R) (2007).

[38] R. Lo Franco, G. Compagno, A. Messina, and A. Napoli, Eur. Phys. J. ST **160**, 247 (2008).

[39] R. Lo Franco, G. Compagno, A. Messina, and A. Napoli, Phys. Lett. A **374**, 2235 (2010).

[40] W. K. Wootters, Phys. Rev. Lett. **80**, 2245 (1998).

[41] L. Henderson, V. Vedral, J. Phys. A **34** 6899 (2001).

[42] X. M. Lu, J. Ma, Z. Xi, and X. Wang, Phys. Rev. A **83**, 012327 (2011).

[43] Q. Chen, C. Zhang, S. Yu, X. X.Yi, and C. H. Oh, Phys.Rev.A **84**,042313 (2011).

[44] Y. Huang, Phys.Rev.A **88**, 014302 (2013).

[45] A. Auyuanet and L. Davidovich, Phys. Rev. A **82**, 032112 (2010).

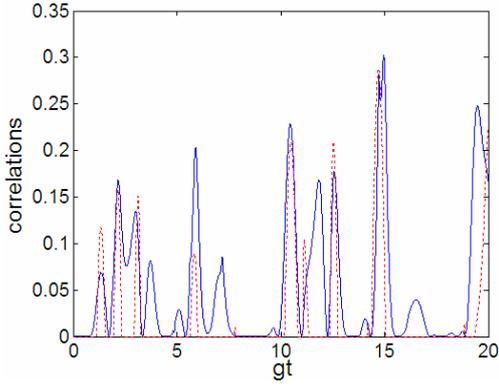

FIG. 1. (Color online) Evolution of the entanglement (red dotted line) and quantum discord (blue solid line) for the initial maximally mixed state of two-qubit $I_4/4$ with the ground state field ($n=0$).

(a)　　　　　　　　　　　　　　　(b)

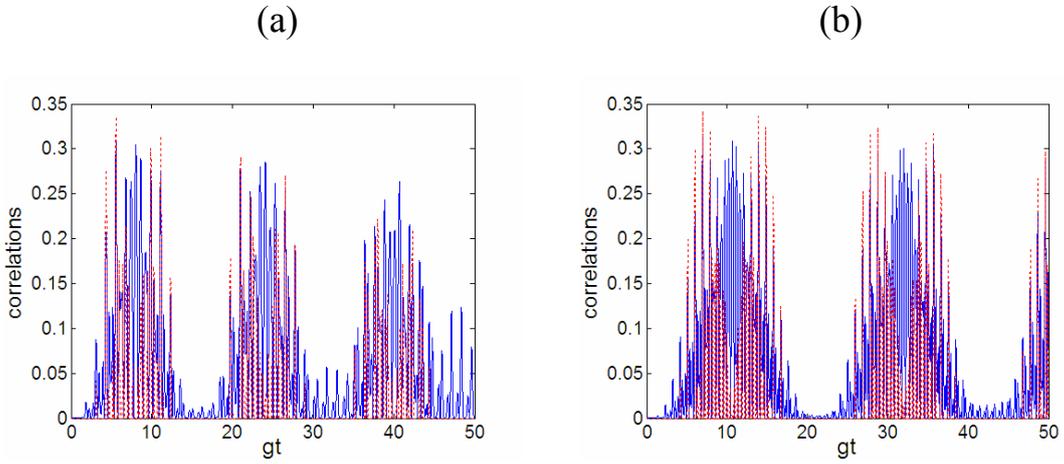

FIG. 2. (Color o nline) The same as figure 1, but ($n=5$) (a) and ($n=10$) (b).

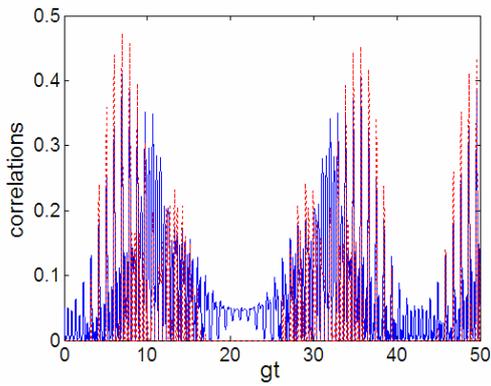

FIG. 3. (Color online) Evolution of the entanglement (red dotted line) quantum discord (blue solid line) for the initial Werner state ($r=0.2$) and Fock state field ($n=10$).